\documentstyle[12pt]{article}
\begin{document}
\baselineskip=0.75cm
\renewcommand{\thesection}{\Roman{section}.}
\renewcommand{\thesubsection}{\Alph{subsection}.}
%
%
\begin{center}
{\large\bf Atomic Radiative Transitions in Thermo Field Dynamics}
\vspace{1.1cm}

J. L. Tomazelli
\vspace{0.5cm}

{\em Departamento de F\'{\i}sica e Qu\'{\i}mica,
Universidade Estadual Paulista, Campus da Guaratinguet\'a,
Av. Dr. Ariberto Pereira da Cunha 333,
12500-000 \\ Guaratinguet\'a, SP, Brazil.}
\vspace{0.8cm}

L. C. Costa
\vspace{0.5cm}

{\em Instituto de F\'{\i}sica Te\'{o}rica,
Universidade Estadual Paulista, \\
01405-900, S\~{a}o Paulo, Brazil.}
\end{center}
\vspace{1.2cm}

\begin{center}
{\bf Abstract}
\end{center}
\vspace{0.5cm}

In this work we rederive the Lamb-Retherford energy shift for an
atomic electron in the presence of a thermal radiation. Using the
Dalibard, Dupont-Roc and Cohen-Tannoudji (DDC) formalism, where
physical observables are expressed as convolutions of suitable
statistical functions, we construct the electromagnetic field
propagator of Thermo Field Dynamics in the Coulomb gauge in order to
investigate finite temperature effects on the atomic energy levels.
In the same context, we also analyze the problem of the ground state
stability. 
%
\newpage
\def\0{\begin{equation}}
\def\1{\end{equation}}
\def\2{\begin{eqnarray}}
\def\3{\end{eqnarray}}
\def\>{{\rangle}}
\def\<{{\langle}}
\def\A{{\bf {A}}}
\def\H{{\bf {H}}}
\def\h{{\it {H}}}
\def\p{{\bf {p}}}
\def\E{{\bf {E}}}
\def\B{{\bf {B}}}
\def\I{{\bf {I}}}
\def\r{{\bf {r}}}
\def\J{{\bf {J}}}
\def\D{{\bf {D}}}
\def\E{{\bf {E}}}
\def\P{{\bf {P}}}
\def\M{{\bf {M}}}
\def\N{{\bf {N}}}
\def\Q{{\bf {Q}}}
\def\I{{\bf {I}}}
\def\V{{\cal {V}}}
\def\k{{\bf {k}}}
\def\e{{\bf {e}}}
%
%
\section{Introduction}
Since the 70's it has been argued \cite{SE73} \cite{MI75} that the
physical interpretation of radiative phenomena, in particular the
shift of atomic energy levels, rely upon different choices in the
ordering of atomic and field operators in the interaction
Hamiltonian.

Almost two decades ago Dalibard, Dupont-Roc and Cohen-Tannoudji
(DDC) \cite{CT82} considered the interaction between a
non-relativistic atomic electron and the quantized electromagnetic
field, showing that the above mentioned arbitrariness can be removed
by requiring the observables´ variation rates to be Hermitian, if we
want them to have a physical meaning. They generalized their
procedure to the case of a small system ${\cal S}$ interacting with
a large reservoir ${\cal R}$ (which may be in thermal equilibrium).
This construct allowed them to separate the physical processes in
two categories, those where ${\cal R}$ fluctuates and polarizes
${\cal S}$ (effects of reservoir fluctuations), and those where
${\cal S}$ polarizes ${\cal R}$ (effects of self-reaction or
radiation reaction).

In the present work we are interested in analyzing the temperature
effects in the context of DDC formalism, where the statistical
functions, which are defined from two-point correlation functions,
play a fundamental role. These functions enable us to obtain
expressions, up to second order in perturbation theory, in terms of
products of symmetrical correlation functions and susceptibilities
\cite{CT98}. The temperature implementation \cite{CT82} can be
made directly in such statistical functions using the equipartition
theorem, leading to a finite temperature description of the relevant
phenomena.

In an alternative way, we shall study the problem using Umezawa's
formalism, known as Thermo Field Dynamics (TFD) \cite{UM96}. In TFD,
the quantum statistical average of a physical observable in a given
ensemble is identified with its expectation value in a thermal
vacuum. In this approach, temperature is introduced as an input in
the eigenstates of the number operator associated to the quantized
field.

Our idea is to investigate the thermal propagator of the
electromagnetic field in the Coulomb gauge in order to identify the
symmetric correlation functions and susceptibilities of DDC
formalism. This is the matter of section III, after a brief presentation
of the main results of DDC construct in section II. In section IV we
investigate the temperature dependence of the Lamb-Retherford energy
shift of an atomic electron in the presence of a thermal radiation
field. In section V we calculate the variation rate of the mean
atomic energy and discuss the stability of the ground state at
finite temperature. In both cases, we are assuming that the whole
system is in thermodynamic equilibrium. Finally, in section VI, we
draw some conclusions.
%
%
\section{Radiation considered as a Reservoir}
In Dalibard, Dupont-Roc and Cohen-Tannoudji \cite{CT82} construct,
the interaction between an atom and the free electromagnetic field
can be seen as the interaction of a microscopic system ${\cal S}$
with a large reservoir ${\cal R}$, in the sense that ${\cal R}$ has
many degrees of freedom and the correlation time among observables
of ${\cal R}$ is small, allowing a perturbative treatment of the
effect due to the coupling of ${\cal S}$ and ${\cal R}$.

Considering ${\cal S}$ an atom fixed at the origin of the coordinate
system and ${\cal R}$ an homogeneous and isotropic broadband
radiation field they addressed, among others, the problem of atomic
radiative corrections as the Lamb shift and the dynamic AC Stark
effect. In particular $\footnote{Here the total hamiltonian is given
by $H = H_S + H_R + H_I$, where $H_S$ describes the atomic system,
$H_R$ the radiation field and $H_I$ is the interaction
Hamiltonian.}$ , they showed that the shift in an atomic energy
level (say $a$) caused by its interaction with the radiation field
can be expressed as
\begin{eqnarray}
(\delta H_{S a})^{\rm fr}=-\frac{1}{2 \hbar}\int_{-\infty}^{\infty}\frac{d\omega}
{2\pi}\hat{\chi}'_{S a}(\omega)\hat{C}_R(\omega)\,\,,\\
(\delta H_{S a})^{\rm rr}=-\frac{1}{2 \hbar}\int_{-\infty}^{\infty}\frac{d\omega}
{2\pi}\hat{\chi}'_{R}(\omega)\hat{C}_{S, a}(\omega)\,\,.
\end{eqnarray}
where $\hat{C}_{S a}(\omega)$ (resp. $\hat{C}_{R}(\omega)$) and
$\hat{\chi}_{S a}(\omega)$ (resp. $\hat{\chi}_{R}(\omega)$) are,
respectively, the symmetric correlation and the even parity part of
the susceptibility functions related to the atomic system (resp.
reservoir) in frequency space. Their true meaning is well
established in the context of DDC formalism which associates (1) to
the reservoir fluctuation effects and (2) to the radiation reaction
effects.

Our main concern is the fact that such statistical functions are
defined from two point functions of the dynamic operators involved
in the interaction Hamiltonian $H_I$. In the case we are interested in,
the interaction Hamiltonian in the Coulomb gauge is given by 
\0 
H_I = - \left(\frac{e}{m}\right) {\bf p} \cdot {\bf A}({\bf 0}), 
\1
where ${\bf p}$ is the momentum associated to the electron's motion
and ${\bf A}({\bf 0})$ is the electromagnetic potential in the
dipole approximation. The two point function for the spatial
component of the field variable $A_i({\bf 0}, t) = A(t)$ 
($i = x, y,z $) is given by 
\0 
g(\tau)=\frac{1}{2} \< \{ A(t'), A(t'') \}_+
\>_A +\frac{i}{2} \< [A(t'), A(t'')/i]_- \>_A\,\,, 
\1 
where $\tau = t'- t''$ and $\<,\>_A$ indicates an average on the reservoir 
state defined by a given statistical weight. As pointed before, the first
term in (4) corresponds to the symmetric correlation function and
the second is related to the linear susceptibility of the reservoir.
The symmetric correlation function of the observable $A(t)$, 
\0
C_R(\tau) = \frac{1}{2} \< \{ A(t'), A(t'') \}_+ \>_A, 
\1 
is real and tends to the ordinary correlation function in the 
classical limit. It gives a physical description of the dynamics 
of fluctuations of the observable $A(t)$. The other statistical
function is the linear susceptibility $\chi_R(\tau)$, which
characterizes the reservoir response to an external perturbation,
defined by 
\2
\chi_R(\tau) &=& \frac{i}{\hbar} \theta(\tau) \< [ A(t'), A(t'') ]_- \>_A \nonumber \\
&=& \frac{2}{\hbar} \theta(\tau) {\rm Im}\,\,g(- \tau),
\3
where $\theta(\tau)$ is the step function.

Since we are interested in analyzing the finite temperature
dependency of (1) and (2), we postpone the calculation of the above
statistical functions for the field components to the next section,
where we shall employ the TFD formalism in order to obtain the
finite temperature two point functions for the radiation field.

Nevertheless, we restrict ourselves to the present action of the
corresponding correlation and susceptibility functions in frequency
space for the $x$ component\footnote{It can be shown that for $i
\neq j$ the statistical functions vanish \cite{LU00}.} of the atomic
variable $( e {\bf p}/m )$, in the situation the atom is found in a
given state $|a\>$ (with $H_S |a\rangle = E_a |a\rangle$), namely 
\2
\hat{C}_{S a}^{xx}(\omega)&=&\sum_b\frac{e^2}{m^2}|\<a|p_x|b\>|^2\pi
[\delta(\omega_{ab}+\omega)+\delta(\omega_{ab}-\omega)], \\
\hat{\chi}_{S a}'^{xx}(\omega)&=&\sum_b\frac{-e^2}{\hbar m^2}|\<a|p_x|b\>|^2
\left[ {\cal P}\frac{1}{\omega_{ab}+\omega}+{\cal P}\frac{1}{\omega_{ab}-\omega}\right], \\
\hat{\chi}_{S a}''^{ xx}(\omega)&=&\sum_b\frac{e^2}{\hbar
m^2}|\<a|p_x|b\>|^2
\pi[\delta(\omega_{ab}+\omega)-\delta(\omega_{ab}-\omega)]. 
\3
Expressions (8) and (9) are obtained by splitting the atomic
susceptibility according to $\hat{\chi}_{S a}(\omega)=
\hat{\chi}'_{S a}(\omega) + i\hat{\chi}''_{S a}(\omega)$, where each
part characterizes, respectively, the response in phase and in
quadrature at the frequency $\omega$. In expression (8), ${\cal P}$
denotes the principal value.
%
%
\section{Thermal Correlation Functions and Susceptibilities}
In this section we study the thermal propagator of the
electromagnetic field of Thermo Field Dynamics (TFD). Our idea is
to define the statistical functions $C_R (\omega)$ and
$\chi_R'(\omega)$ from the appropriated  propagator of QED,
implementing temperature at the beginning. We start by writing the
quantized electromagnetic potential $A_{\;i} (t)$ as\footnote{As in
the last section, we assume that the atom is at rest at the origin
of the coordinate system (${\bf r=0}$) and that we are in the dipole
approximation.} 
\0 
A_{\;i} (t) = A_{\;i}^{(+)} (t) + A_{\;i}^{(-)}
(t), 
\1 
where $A_{\;i}^{(+)} (t)$ and $A_{\;i}^{(-)} (t)$ are the
components with positive and negative frequency, defined
respectively as 
\0 
A_{\;i}^{(+)} (t) = \sum_{\k,r} \alpha_\k \;
\e_i^r (\k)  \; a_{\k}^r \; e^{- i \omega_\k t}, \1 \0 A_{\;i}^{(-)}
(t) = \sum_{\k,r} \alpha_\k \; \e_i^r (\k)  \; {a_{\k}^r}^\dagger \;
e^{ i \omega_\k t}, 
\1 
with 
\0 \alpha_{\k} = { \left(
{\frac{\hbar}{2 \varepsilon_0 L^3 \omega_\k}} \right) }^{1/2}. 
\1

In TFD we double the field degrees of freedom introducing the tilde conjugated
of $A_{\;i} (t)$ \cite{UM96} \cite{UM95}. Using the thermal doublet notation 
\cite{UM95} \cite{MA85}, we obtain
\0
{\A_{\;i} (t)}
        =
        \left(
        \begin{array}{c}
        A_{\;i} (t) \\
        \tilde{A}_{\;i} (t)
        \end{array}
        \right) \;\;\;\;\;\;\;\; \bar{\A}_{\;i} (t) =
( \; {A}_{\;i} (t), \; - \; \tilde{A}_{\;i} (t) \; ),
\1
where
($\;\bar{} \;$) denotes the transposed and 
\2 
A_{\;i}  (t) &=&
\sum_{\k,r} \alpha_\k \; \e_i^r (\k)  \;(\; a_{\k}^r
\; e^{- i \omega_\k t}  +  {a_{\k}^r}^\dagger  \; e^{ i \omega_\k t} \; ) \nonumber \\
&=& A_{\;i}^{(+)} (t) + A_{\;i}^{(-)}  (t) ,
\3
\2
\tilde{A}_{\;i}  (t) &=& \sum_{\k,r} \alpha_\k \; \e_i^r (\k)
\; ( \; \tilde{a}_{\k}^r \;  e^{ i \omega_\k t}  +  {\tilde{a}_{\k}^r}{}^{\dagger}
\; e^{ - i \omega_\k t} \; ) \nonumber \\
&=&  \tilde{A}_{\;i}^{(+)} (t) + \tilde{A}_{\;i}^{(-)} (t). 
\3 
By
construction, both fields $A_{\;i}$ and $\tilde{A}_{\;i}$ are
independent; the corresponding absorption and emission operators
satisfy the algebra \cite{UM95} 
\0 
[ \; a_{\k}^r,
{a_{\k'}^s}{}^\dagger \; ] = [\; \tilde{a}_{\k}^r ,
{\tilde{a}_{\k}^s}{}^{\dagger} \;] = \delta_{\k,\k'} \;
\delta_{r,s}. 
\1 
At zero temperature, the vacuum state is given by
the direct product $|0\>_{A} \; \otimes \; |0\>_{\tilde{A}} \;
\dot{=} \; |0\>$. Using (17), it follows that 
\0 
A_{\;i}^{(+)} \;
|0\> = 0, \;\;\;\;\;\; \tilde{A}_{\;i}^{(+)} \; |0\> = 0. 
\1 
In order to find the thermal propagator associated with the statistical
functions, we must calculate the commutator 
\0 
[ \A_i (t'),
\bar{\A}_j (t'') ]^{\mu \nu}  \; = \; \Delta_{i j}^{\mu \nu} (\tau),
\1 
where $\mu, \nu$ = 1,2 and $i,j$ = $x,y,z$. The anti-diagonal
components of the above quantity are identically zero when their
expectation value in the $|0\>$ state is taken. The component $\mu =
\nu = 1$ can be written as 
\0 
\Delta_{\;ij}^{11}(\tau) =
\Delta_{\;ij}^{11 \;\; (+)}(\tau) + \Delta_{\;ij}^{11 \;\;
(-)}(\tau), \1 where \0 {\Delta_{\;ij}^{11}}^{(+)} (\tau) \; \dot{=}
\; [ \; A_{\;i}^{(+)} (t'), A_{\;j}^{(-)} (t'') \; ], \1 \0
{\Delta_{\;ij}^{11}}^{(-)} (\tau) \; \dot{=} \; [ \; A_{\;i}^{(-)}
(t'), A_{\;j}^{(+)} (t'') \; ]. 
\1 
Now, using (11), (12), (17) and (18), we obtain, 
\0 
{\Delta_{\;ij}^{11}}^{(+)} (\tau) \; =
\sum_{\k,r} \; \alpha_{\k}^2 \; \e_i^r (\k) \; \e_j^r (\k) \; e^{- i
\omega_\k \tau}, 
\1 
\0 
{\Delta_{\;ij}^{11}}^{(-)} (\tau) \; = -
\sum_{\k,r} \; \alpha_{\k}^2 \; \e_i^r (\k) \; \e_j^r (\k) \; e^{ i
\omega_\k \tau}. 
\1 
From (23) and (24), we can define two functionals: 
\2 
{\Delta_{\;ij}^{11}}_{(ret)} (\tau) \; &{\dot{=}}&
\; \theta (\tau) {\Delta_{\;i j}^{11}}^{(+)} (\tau) +
\theta (\tau) {\Delta_{\;i j}^{11}}^{(-)} (\tau) \nonumber \\
&=& {\Delta_{\;i j \; (ret)}^{11}}^{(+)} (\tau) +
{\Delta_{\;ij\;(ret)}^{11}}^{(-)} (\tau), 
\3 
and 
\0
{\Delta_{\;ij\;(1)}^{11}} (\tau) \; \dot{=} \;
{\Delta_{\;ij}^{11}}^{(+)} (\tau) - {\Delta_{\;ij}^{11}}^{(-)}
(\tau). 
\1 
It can be easily shown that, at zero temperature, 
\0
{\Delta_{\;ij\;(1)}^{11}} (\tau) \; = \< 0 | \{ A_i (t'), A_j (t'')\} | 0 \> \;. 
\1 
So, we should point out that the quantity
$g(\tau)$, defined before as a two time average of a given
observable, is associated in the present case with the functional 
\0
g_{ij} (\tau)=  \< 0 | A_i (t') A_j (t'') | 0 \> = \frac{1}{2}
\Delta_{\;ij\;(1)}^{11} (\tau) + \frac{1}{2} \Delta_{\;ij}^{11}
(\tau) \; , 
\1 
according to expression (4).

By taking the Fourier transform of (25) and (26) we obtain, respectively,
\2
{\Delta_{ij\;(ret)}^{11}}(\omega)&=& \sum_{\k,r}\alpha_{\k}^2 \; \e_i^r (\k) \; \e_j^r (\k)
\left[\left(\frac{i}{\omega-\omega_\k+i\epsilon}\right)
-\left(\frac{i}{\omega+\omega_\k+i\epsilon}\right)\right]. \\
{\Delta_{ij\;(1)}^{11}}(\omega)&=&\sum_{\k,r}\alpha_{\k}^2 \; \e_i^r
(\k) \; \e_j^r (\k)\pi [\delta(\omega + \omega_\k) + \delta(\omega -
\omega_\k)]. 
\3 
Adopting the same procedure, we can extend the above
calculation to the component $\mu = \nu = 2$. As a result, we have
\2 
{\Delta_{ij(ret)}^{22}}(\omega)&=&\sum_{\k,r} \alpha_{\k}^2
\e_i^r (\k) \e_j^r (\k)
\left[\left(\frac{i}{\omega-\omega_\k-i\epsilon}\right)-
\left(\frac{i}{\omega+\omega_\k-i\epsilon}\right)\right], \\
{\Delta_{ij\;(1)}^{22}}(\omega)&=&-\sum_{\k,r} \; \alpha_{\k}^2
\e_i^r(\k) \e_j^r(\k) \pi [
\delta(\omega+\omega_\k)+\delta(\omega-\omega_\k) ]. 
\3 
We may write expressions (29) and (31) in a more compact notation, i.e., 
\2
\Delta_{ij\;(ret)}(\omega) = \sum_{\k,r}\alpha_{\k}^2
\e_i^r(\k)\e_j^r(\k)\left\{\frac{i}{k_0-\omega_\k+i{\bf
\tau_3}\epsilon}-\frac{i}{k_0+ \omega_\k+i{\bf
\tau_3}\epsilon}\right\} 
\3 
and, in the same way, we write (30) and (32) as 
\2 
\Delta_{ij\;(1)}(\omega)=-\sum_{\k,r}\alpha_{\k}^2\e_i^r
(\k)\e_j^r (\k) \pi {\bf \tau_3}[
\delta(\omega+\omega_\k)+\delta(\omega-\omega_\k) ], 
\3 
where, in the last two expressions, 
\0 
{\bf \tau_3} = \left(
\begin{array}{cc}
1 & 0 \\
0 & -1
\end{array} \right).
\1
In TFD, it is known that the propagator at zero temperature is related to the one
calculated in the thermal vacuum through a Bogoliubov transformation \cite{MA77}.
Applying this result to (33) and (34), we obtain, respectively,
\2
\Delta_{\;i j (ret)}^{ \; \mu \nu \; \beta } (\omega) &=&  \{ B_{\k}^{-1} (\beta) \;
\Delta_{\;i j \; (ret)} (\omega) \; B_{\k} (\beta) \}^{\mu \nu}, \\
\Delta_{\;i j \; (1)}^{ \; \mu \nu  \; \beta } (\omega)  &=& \{ B_{\k}^{-1} (\beta) \;
\Delta_{\;i j \; (1)} (\omega) \; B_{\k} (\beta) \}^{\mu \nu},
\3
where $B_\k (\beta)$ is give by
\0
B_\k (\beta) = (1 - n_\k )^{1/2}
        \left(
        \begin{array}{cc}
        1 & - \; {f_\k}^{\alpha} \\
        - \; {f_\k}^{ 1 - \alpha } & 1
        \end{array}
        \right),
\1 with $\alpha = 1/2$, $f_\k = {\rm exp}[- \hbar \omega_\k \beta]$ and 
\0 
n_\k = \frac{1}{f_\k^{-1} -1} = \frac{1}{e^{\hbar \omega_\k \beta} - 1}, 
\1 
($\beta = 1/ kT$, where $k$ is the Boltzmann constant and $T$ the equilibrium temperature).

The $\mu = \nu = 1$ component of (36) is found to be
\2
\Delta_{\;i j \; (ret)}^{11 \;\; \beta} (\omega)  &=& - i \sum_{\k,r} \; \alpha_{\k}^2 \;
\e_i^r (\k) \; \e_j^r (\k) \;  \Big\{ \;  {\cal{P}} \frac{1}{\omega_\k - \omega} \; + \;
{\cal{P}} \frac{1}{\omega_\k + \omega } \; + \nonumber \\
 &+&  i \; \pi \; [ \; \delta(\omega_\k - \omega) - \delta(\omega_\k + \omega) \; ] \;
 (1 + 2 n (\omega_\k) ) \;\Big\},
\3 
and, from (37), 
\0 
\Delta_{\;i j \; (1)}^{11 \;\; \beta} (\omega)
=  \sum_{\k,r} \; \alpha_{\k}^2 \; \e_i^r (\k) \; \e_j^r (\k) \; \pi
\; [ \; \delta(\omega - \omega_\k) + \delta(\omega + \omega_\k) \; ]
\; (1 + 2 n (\omega_\k) ). 
\1 
Now, relating (5) to (41) and (6) to (40), we are in position to define the 
thermal correlation function and susceptibilities, 
\0 
C_{\;i j}^{\; \beta} (\omega) \;
\dot{=} \; \Delta_{\;i j \; (1)}^{11 \;\; \beta} (\omega), 
\1 
and 
\0
\chi_{\;i j}^{\;\beta} (\omega) \; \dot{=} \; \frac{i}{\hbar} \;
\Delta_{\;i j \; (ret)}^{11 \;\; \beta} (\omega), 
\1 
where, again, we split (43) as 
\0 
\chi_{\;i j}^{\; \beta} (\omega) = \chi_{\;i
j}^{ \; '\; \beta } (\omega) + i \; \chi_{\;i j}^{\; '' \; \beta }
(\omega), \1 \0 \chi_{\;i j}^{\; ' \; \beta } (\omega) =
\frac{1}{\hbar} \; \sum_{\k,r} \; \alpha_{\k}^2 \; \e_i^r (\k) \;
\e_j^r (\k) \;  \left[ \; {\cal{P}} \frac{1}{\omega_\k - \omega} \;
+ \; {\cal{P}} \frac{1}{\omega_\k + \omega } \;\right], 
\1 
\0
\chi_{\;i j}^{\; '' \; \beta } (\omega) = - \frac{1}{\hbar} \;
\sum_{\k,r} \; \alpha_{\k}^2 \; \e_i^r (\k) \; \e_j^r (\k) \; \pi \;
(1 + 2 n (\omega_\k) ) \; [ \; \delta(\omega_\k + \omega) -
\delta(\omega_\k - \omega) \; ]. 
\1 
Choosing $i=j=x$ and substituting the summation over modes by a polarization sum and an
integral in $\k$, we obtain 
\2 
C_{\;x x}^{\; \beta} (\omega) \; &=&
\; \frac{1}{3\pi\varepsilon_0c^3} \int_0^{\omega_M} d\omega' \;
\hbar \omega' ( n(\omega') + 1/2 ) \;
[ \; \delta(\omega' - \omega) + \delta(\omega' + \omega) \; ] \nonumber \\
&=& \frac{1}{3\pi\varepsilon_0c^3} \; \hbar|\omega| ( n(|\omega|)+ 1/2 ) \\
\chi_{\;x x}^{\; ' \; \beta } (\omega) &=&  \; \frac{1}{6 \pi \varepsilon_0 c^3 }
\int_0^{\omega_M} d\omega'  \omega' \left[ \; {\cal{P}}
\frac{1}{\omega' - \omega} \; + \; {\cal{P}} \frac{1}{\omega' + \omega } \;\right] \\
\chi_{\;x x}^{\; '' \; \beta } (\omega) &=& \frac{- 1}{6 \pi \varepsilon_0 c^3 }
\int_0^{\omega_M} d\omega' \omega' \; ( 2 n (\omega') + 1 ) \;
[ \; \delta(\omega' + \omega) - \delta(\omega' - \omega) \; ] \nonumber \\
&=& \frac{1}{3\pi\varepsilon_0c^3} \; \omega \; ( n(\vert \omega \vert)+ 1/2 ).
\3


\section{The Lamb-Retherford Shift via TFD}

We now apply the above results to the case of Lamb-Retherford shift and discuss
the related thermal radiative effects. As mentioned in section I, this is done by
substituting (8) and (47) into expression (1) which, according to \cite{CT98},
represent the desirable radiative correction,
\begin{flushleft}
$\displaystyle{\hbar(\delta H_{S a})_{\beta}^{\rm fr}=\frac{e^2}{6\pi^2\varepsilon_0m^2c^3}
\sum_b|\<a|{\bf p}|b\>|^2\times}$
\end{flushleft}
\begin{equation}
\times\int_0^{\omega_M}d\omega'\omega'\<n(\omega')+\frac{1}{2}\>\left[
{\cal P}\frac{1}{\omega_{ab}+\omega'}+{\cal P}\frac{1}{\omega_{ab}-\omega'}
\right]\,\,.
\end{equation}
The atomic energy shift due to the field fluctuations appears as a
sum of the effects $(\delta H_{S a}){}^{\rm fr '}$ of the ``thermal
photons'', proportional to $\<n(\omega)\>$, and that of the vacuum
fluctuations $(\delta H_{S a})^{\rm fv}$, corresponding to the ``$
\hbar \omega / 2$ by mode''. This last term can be manipulated using
the relations 
\0 
\int_0^{\omega_M}  \omega' \; d \omega' \; {\cal P}
\frac{1}{\omega' \pm \omega_0} = \omega_M \mp \omega_0 \; {\rm ln}
\frac{\omega_M}{\omega_0} + {\cal O} \Big( \frac{\omega_0}{\omega_M}
\Big) \1 \0 \int_0^{\omega_M}  \omega' \; d \omega' \left[  {\cal P}
\frac{1}{\omega' + \omega_0} + {\cal P} \frac{1}{\omega_0 - \omega'}
\right] = - 2 \omega_0 \; {\rm ln} \frac{\omega_M}{\omega_0}. 
\1
Hence, we obtain 
\0 \hbar (\delta H_a)^{\rm fv} = \frac{e^2}{6 \pi^2
\varepsilon_0 m^2 c^3} \; \sum_b |\<a|{\bf p}|b\>|^2 \; (-
\omega_{ab} ) \; {\rm ln} \frac{\omega_M}{\vert \omega_{ab} \vert},
\1 or \0 \hbar (\delta H_a)^{\rm fv} = \frac{\alpha}{3 \pi} \Big(
\frac{\hbar}{m c} \Big)^2 \Big( {\rm ln} \frac{\omega_M}{c K_a}
\Big) \<a| \frac{e^2}{\varepsilon_0} \delta(\r) |a\>, 
\1 
where $\alpha $ is the fine structure constant and $\hbar c K $ is the
mean atomic excitation energy. Expression (54) corresponds to the
(pure) Lamb-Retherford shift as found in literature \cite{MA94}. It 
is well known that its physical origin comes from the vacuum fluctuation 
of the radiation field (reservoir). The presence of $\hbar$ in (54) shows
the quantum character of this effect, just as the vacuum fluctuation
which gives rise to it.

The contribution $(\delta H_{S a}){}^{\rm fr '}$ proportional to
$\<n(\omega')\>$ correspond to a stimulated radiative correction due
to the ``thermal photons''. It resembles the AC Stark effect when
the thermal radiation field is substituted by a quantized
electromagnetic field. In the present context $(\delta H_{S
a}){}^{\rm fr '}$ corresponds to thermal radiative correction to the
(pure) Lamb-Retherford shift and its effect vanishes as the temperature 
approach to zero.
%
%
\section{Energy Exchange}
In order to analyze the effects of the thermal reservoir on the
stability of the atomic ground state, we now consider the energy
exchange between a bound electron and the thermal radiation field
using the results of section III.  Following \cite{CT98}, the
variation rate of the mean atomic energy when the system
is in its ground state (say $a$) is given by 
\0 
\frac{d}{dt} \< H_S
\>_{a}^{\beta} = \sum_b (E_b - E_a) \Gamma_{a \rightarrow b}, 
\1
where $\Gamma_{a \rightarrow b}$ represents the transition rate
between the ground state $a$ and an excited state $b$ due to the
interaction with the reservoir. It is shown in reference \cite{CT82}
that (55) can be written as 
\0 
\frac{d}{dt} \< H_S \>_{a}^{\beta} =
{\dot {\cal Q}}_{\beta}^{\rm fr} + {\dot {\cal Q}}_{\beta}^{\rm rr},
\1 
where 
\2 {\dot {\cal Q}}_{\beta}^{\rm fr} &=& \int \frac{d
\omega}{2 \pi} \; \omega \;
{\hat C}_{R}^{\beta} (\omega) {\hat \chi}_{Sa}'' (\omega), \\
{\dot {\cal Q}}_{\beta}^{\rm rr} &=& - \int \frac{d \omega}{2 \pi} \; \omega \;
{{\hat \chi}_{R}^{\beta ''}} (\omega) {\hat C}_{Sa} (\omega).
\3
The last two expressions have a clear meaning: (57) is associated with the energy 
absorption by the system when it is affected by reservoir fluctuations and (58) is 
related to the damping of the atomic motion caused by the reservoir.

Using expressions (9) and (47) and taking into account the spatial
components $x$, $y$ and $z$ of the electromagnetic potential, we
find that (57) can be written as 
\2
{\dot {\cal Q}}_{\beta}^{\rm fr} &=& {\dot {\cal Q}}^{\rm fr '} 
+ {\dot {\cal Q}}^{\rm fv} \nonumber \\
&=& \sum_b (E_b - E_a) \Gamma_{ab}^{sp} [ \<n(\vert\omega_{ab}\vert)\> + 1/2],
\3 
where 
\0 
\Gamma_{ab}^{sp} = \frac{e^2 \vert \< a|\p|b\> \vert^2
\vert \omega_{ab} \vert} {3 \pi \varepsilon_0 \hbar m^2c^3} 
\1 
is the rate of spontaneous emission related to the transition between
the levels $b$ and $a$.

The quantity ${\dot {\cal Q}}_{\beta}^{\rm rr}$ is calculated in the same way from expressions
(7), (49) e (58). As a result, we find
\2
{\dot {\cal Q}}_{\beta}^{\rm rr} &=& {\dot {\cal Q}}^{\rm fr '} + {\dot {\cal Q}}^{\rm fv} \nonumber \\
&=& - \sum_b (E_b - E_a) \Gamma_{ab}^{sp} [ \<n(\vert\omega_{ab}\vert)\> + 1/2 ].
\3 
Substituting (59) and (61) in (56) we conclude that 
\0
\frac{d}{dt} \< H_S \>_a^{\beta} = 0. 
\1 
This result is what we must expect since the whole system is in thermal 
equilibrium at temperature $T$. In the present context, one can say that, 
in thermodynamic equilibrium, the bound electron reaches a new ground
state which corresponds to the original one shifted by the amount
$(\delta H_{S a})_{\beta}^{\rm fr}$. For $T=0$, the ground state
stability still holds, since the effects of radiation reaction,
${\dot {\cal Q}}_{\infty}^{\rm rr}$, are cancelled by the effects of
thermal vacuum fluctuations, ${\dot {\cal Q}}_{\infty}^{\rm fv}$.
%
%
\section{Concluding Remarks}
In the present work we have used the structure of DDC construct to
implement temperature effects via TFD. After a brief review of the 
main DDC results, we have investigated the propagators of the
electromagnetic field in the context of TFD and derived the
symmetric correlation functions and susceptibilities for the field
variable $A({\bf 0}, t)$. Applying the results to the case of an
atomic electron interacting with a thermal radiation field, we
calculate the Lamb-Retherford energy shift and the corresponding
corrections due to thermal photons.

In the last section we have analyzed the energy exchange between the
atomic electron and the thermal radiation field and concluded that,
once the whole system is in thermodynamic equilibrium at a given
temperature $T$, the stability of the ground state is maintained,
even when $T$ approaches to zero.

We must point out that the original DDC formalism includes the case
where the reservoir is a thermal radiation field. As remarked in
\cite{CT82}, this is done by replacing the mean number of particles
($n ( \vert \omega \vert)$) by a Bose-Einsten distribution in the
resulting statistical functions. However, such procedure differs
from ours in the sense that the detailed balance principle become
meaningless in the context of TFD where the population dynamics
between two given atomic states is not accessible.

Finally, we mention that the applicability of TFD in the scope of
DDC formalism is not restricted to the problem we have just
revisited. Among the physical phenomena we intend to investigate in
the near future are those related to the dissipative processes in
quantum optics \cite{FU98} \cite{CA83}.
%
%
\vspace{0.5cm}

\noindent{\bf Acknowledgements.} JLT acknowledges CNPq for partial support and IFT/UNESP
for the hospitality. LCC is grateful to FAPESP for the financial support. The authors would 
like to thank professor H. M. Fran\c ca for helpful suggestions. We also aknowledge the 
Referee for the careful reading of the manuscript.
%
%


\newpage
\begin{thebibliography}{99}
%
\bibitem{SE73}
I. R. Senitzky, Phys. Rev. Lett. {\bf 31}, 955 (1973).
%
\bibitem{MI75}
P. W. Milonni and W. A. Smith, Phys. Rev. A {\bf 11}, 814 (1975).
%
\bibitem{CT82}
J. Dalibard, J. Dupont-Roc and C. Cohen-Tannoudji, J. de Physique {\bf 43}, 1617 (1982), J.
de Physique {\bf45}, 637 (1984).
%
\bibitem{CT98}
 C. Cohen-Tannoudji, J. Dupont-Roc and G. Grynberg,
``Atom-Photon Interactions - Basic Processes and Applications'', J. Wiley, NY (1998).
%
\bibitem{UM96}
H. Umezawa and Y. Takahashi, Int. J. Mod. Phys. B {\bf 10}, 1755 (1996).
%
\bibitem{UM95}
H. Umezawa, ``Advanced Field Theory'', AIP Press, NY (1995).
%
\bibitem{MA85}
H. Matsumoto, Y. Nakano and H. Umezawa, Phys. Rev. D {\bf 31}, 429 (1985).
%
\bibitem{MA77}
H. Matsumoto, Fortsh. Phys. {\bf 25}, 1 (1977).
%
\bibitem{LU00}
L. C. Costa, master thesis, Instituto de F\'{\i}sica Te\'orica - UNESP, S\~ao Paulo -
Brazil (2000).
%
\bibitem{MA94}
F. Mandl and G. Shaw, Quantum Field Theory, Revised Edition, J. Wiley, (1994).
%
\bibitem{FU98}
K. Fujikawa, Phys. Rev. E {\bf 57}, 5023 (1998) and {\it ibid} {\bf 58}, 7063 (1998).
%
\bibitem{CA83}
A. O. Caldeira and A. J. Leggett, Ann. Phys. {\bf 149}, 374 (1983).
%
%
\end{thebibliography}
\end{document}